\documentclass[aps,prb,twocolumn,showpacs,showkeys,footinbib,superscriptaddress]{revtex4-2}

\usepackage{amsmath}
\usepackage{amssymb}
\usepackage{graphicx}
\usepackage{bm}
\usepackage{color}
\usepackage{relsize}
\usepackage{braket}
\usepackage[caption=false]{subfig}

\usepackage{hyperref}
\usepackage[all]{hypcap}

\begin{document}
\title{ Excitons  and trions in monolayer semiconductors with correlated electrons}
\date{\today}

\author{Dinh Van Tuan}
\email[]{vdinh@ur.rochester.edu}
\affiliation{Department of Electrical and Computer Engineering, University of Rochester, Rochester, New York 14627, USA}
\author{Hanan~Dery}
\email[]{hanan.dery@rochester.edu}
\affiliation{Department of Electrical and Computer Engineering, University of Rochester, Rochester, New York 14627, USA}
\affiliation{Department of Physics and Astronomy, University of Rochester, Rochester, New York 14627, USA}

\begin{abstract} 
We revisit low-temperature optical spectra of transition-metal dichalcogenide monolayers and point to a possible crystallization of electrons (or holes) at low to moderate charge densities. To calculate the excitonic spectra under such conditions, we introduce the recursion method and compute how the charge density affects the energies, linewidths, and oscillator strengths of exciton and trion complexes. Equally important, we study how excitons and trions in the monolayer evolve when the charge particles gradually transition to a periodic Wigner lattice. The results provide valuable information on the ability to detect whether the particles are ordered through the exciton spectrum. Finally, we calculate the change in exciton energy  in cases that the added charge particles have similar and dissimilar  quantum numbers (spin and valley) to those of the electron or hole in the exciton. The results of this work shed new light on important optical properties of monolayer semiconductors.
\end{abstract}


\maketitle

\section{Introduction}

Charge-neutral excitons in semiconductors comprise an electron in the conduction band that is bound to a hole in the valence band \cite{Elliott_PR57, Laude_PR71, RashbaBook, HaugBook, Combescot_Book}. The exciton typically manifests as a discrete resonance in the optical spectrum below the band-gap energy of the semiconductor. A charged exciton emerges as an additional resonance, further below the energy of the neutral exciton, when electrons or holes are added to the semiconductor \cite{Stebe_PRL89, Kheng_PRL93, Finkelstein_PRL95,Huard_PRL00, Bracker_PRB05, Astakhov_PRB00,Wang_RMP18}. The charged exciton can be viewed as a three-body bound complex consisting of two electrons and one hole (negative trion) or one electron and two holes (positive trion). 


There is a plethora of evidence to support the idea that the three particles of a trion are bound and `touch' each other in transition-metal dichalcogenides (TMDs) monolayers at low charge densities.  For example, the trion picture explains how the $g$-factor of each optical transition depends on the valleys and spins of the involved states in the conduction and valence bands \cite{Robert_PRL21,Liu_PRL20b,He_NatComm20}. Similarly, one can explain whether the circular polarization of the emitted light from optical transitions of each of the trions should be co-polarized, cross-polarized, or unpolarized with respect to the polarized laser \cite{He_NatComm20}. The trion picture explains the fine structure of bright negative trions through the exchange interactions between its three particles \cite{Jones_NatPhys16,Hichri_PRB20}. In addition, the energy position, sign and amplitude of the $g$-factor, and whether the emitted light should be cross or co-polarized, can be explained for a multitude of phonon-assisted optical transitions of dark trion species \cite{Yang_PRB22,Liu_PRL20b,He_NatComm20}. 

As the charge density in the monolayer continues to increase, the exciton and trion resonances start to broaden, shift in energy, and their oscillator strengths change \cite{Wang_NanoLett17,Courtade_PRB17,Smolenski_PRL19, Wang_PRX20, Liu_PRL20, Liu_NatComm21, Li_NanoLett22}. In an effort to explain these observations, two main theoretical concepts have been suggested as alternatives to the one with excitons and trions. The first one is the Bronold-Suris theory in which the trion develops into a four-particle complex (tetron), wherein the trion moves together and is correlated with the Coulomb hole in the Fermi sea that is created when the trion forms \cite{Bronold_PRB00,Suris_PSS01,Esser_pssb01,Koudinov_PRL14}. Recently, we have further generalized this idea to composite excitonic states, such as six- and eight-particle complexes (hexcitons and oxcitons), that emerge at large charge densities if the valley-spin configuration is rich enough \cite{VanTuan_PRL22,VanTuan_PRB22}. The second concept is the Fermi-polaron theory, in which the exciton and trion resonances are regarded as repulsive and attractive branches, respectively, of the collective Fermi-sea response to a photoexcited electron-hole pair \cite{Sidler_NatPhys17, Efimkin_PRB17, Efimkin_PRB18, Fey_PRB20}.   Relationships between the exciton-trion and Fermi polaron pictures, or between the latter and Bronold-Suris tetrons have been discussed in recent literature \cite{Rana_PRB20, Efimkin_PRB21, Glazov_JCP20,Chang_PRB18}. 

 \begin{figure*} 
\centering
\includegraphics[width=15cm]{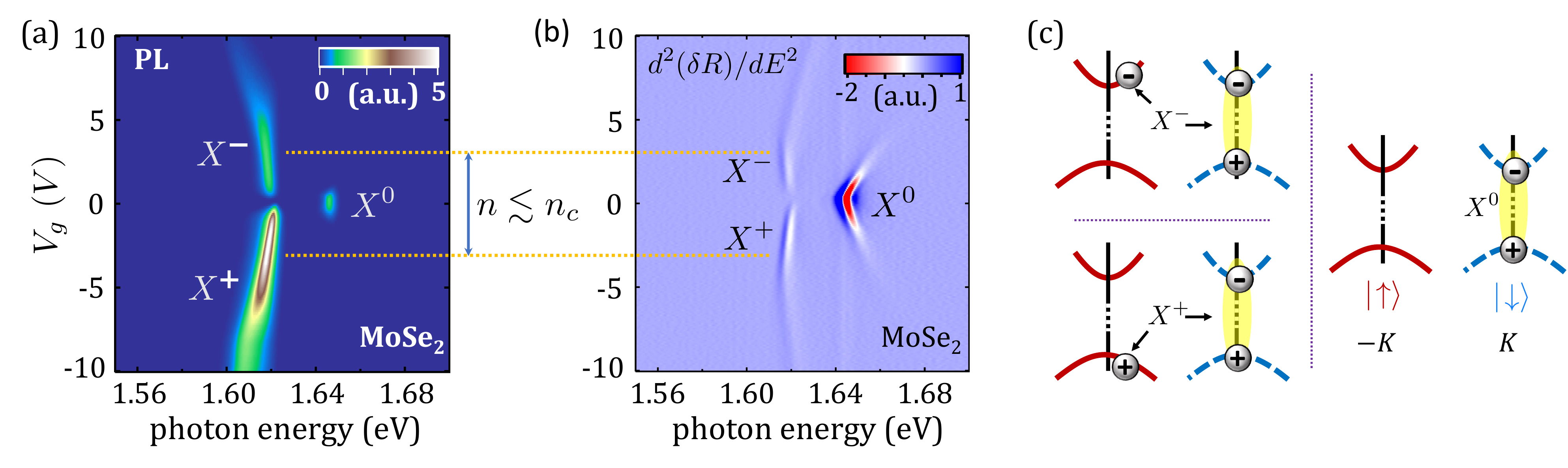}
\caption{(a) and (b) Photoluminescence and differential reflectance spectra of a charge tunable MoSe$_2$ monolayer at 15$\,$K \cite{Liu_NatComm21}.  The resonances are of the ground state neutral exciton ($X^0$), positive trion ($X^+$), and negative trion ($X^-$). Their state compositions are shown in (c) following photoexcitation of the valley at $K$ (highlighted by the yellow ovals).}\label{fig:expMoSe2} 
\end{figure*}

Common to all of these theories is the assumption of continuous translation invariance symmetry, where the wavevector $\mathbf{k}$ of a charge particle is a good quantum number. 
The problem with this assumption is that the translation symmetry could be broken at low temperature and small charge densities, wherein the Coulomb energy of the system exceeds its kinetic energy \cite{Wigner_PR34,Giuliani_Vignale_Book,Attaccalite_PRL02,Hossain_NatPhys21,Smolenski_Nat21,Shimazaki_PRX21,Cassella_PRL23}. For example, diffusion Monte Carlo methods show a transition between a two-dimensional system of itinerant particles to a system of strongly localized particles in a periodic arrangement (Wigner crystal) when $r_s = 1/ \sqrt{\pi a_B^2 n}  \gtrsim 36$ \cite{Attaccalite_PRL02}.  $n$ is the charge density and $a_B = a_0\epsilon/m^\ast$ is the effective Bohr radius, where $a_0 =0.53 ~\AA$,  $\epsilon$  is the relative permittivity of the encapsulating layers, and  $m^\ast$ is the ratio between the masses of a charge particle in the monolayer and free electron in vacuum. Furthermore, recent Fermionic neural network calculations have shown that the order parameter of charge particles rises sharply from zero at  $r_s = 1$ to a finite value already at $r_s = 2$, and the crystalline order continues to grow as $r_s$ increases further \cite{Cassella_PRL23}. A weak photoexcitation of the semiconductor should neither perturb nor polarize the order of the charge particles. The order is established by long-range Coulomb correlations between the particles. On the other hand,  a strong binding between the electron and hole of the exciton (i.e., small-radius exciton) means that the interaction between a charge-neutral exciton and a charge particle has a much shorter range than the lattice constant of the Wigner crystal.

All in all, it is inherently problematic to assume that when the charge density starts to grow in  a low-temperature monolayer semiconductor, the picture of excitons and trions evolves directly to a system wherein the photoexcited pair interacts with itinerant and weakly correlated charge particles. A suitable framework to study this problem is that the photoexcited electron-hole pair interacts with charge particles that are localized or quasi-localized in some fashion, where the localization is introduced by Coulomb correlations rather than disorder. The goal of this work is to address this problem and show how the exciton and trion picture at vanishing charge densities evolves when charge particles are gradually added to the monolayer. The theory we present provides important metrics that help us understand how to identify the order of charge particles from the optical spectrum. Furthermore, the theory elucidates what happens to the optical spectrum when the charge particles gradually lose their ordered state  (e.g., by raising temperature). The results we present successfully reproduce experimental findings regarding the broadening of the optical resonances, their energy shifts, and the evolution of their oscillator strengths when the charge density increases. In addition, we explain how the energy blueshift of the exciton is affected by the distinguishability of its electron and hole components with respect to the charge particles in the monolayer.


This paper is organized as follows. Section~\ref{sec:mot} includes a brief experimental review of excitonic complexes in MoSe$_2$ and WSe$_2$ monolayers. We revisit their low-temperature optical spectra and point to indications of possible crystallization of electrons (or holes) at low to moderate charge densities.  By highlighting relevant experimental findings, we explain the motivation  and importance of our study.  The theory behind the recursion method is presented in Sec.~\ref{sec:theory}. We explain how to apply this method in the  study excitons and trions in monolayer semiconductors. Simulation results and their analyses are presented in Sec.~\ref{sec:results}. A summary and outlook are given in Sec.~\ref{sec:con}. The appendices include technical details on the recursion method and computation procedure.

\section{Background and Motivation} \label{sec:mot}

We analyze the absorption and emission spectra of high-quality MoSe$_2$ and WSe$_2$ monolayers. By pointing to interesting observations, the need for the theory we introduce later in this work becomes apparent. The experimental data we present was provided by courtesy of Chun Hung Lui. Details on device fabrication and optical measurements are found in the published works of Erfu Liu \textit{et al.} in Refs.~\cite{Liu_NatComm21,Liu_PRL20}. Figures~\ref{fig:expMoSe2}(a) and (b) show the photoluminescence and differential reflectance spectra of a charge tunable MoSe$_2$ monolayer at 15$\,$K, respectively. Changing the gate voltage by 1$\,$V in this device amounts to a change of $\sim6.6\times10^{11}$~cm$^{-2}$ in charge density \cite{Liu_NatComm21}. The emission and absorption spectra include the bright exciton resonance at small gate voltages ($X^0$), the positive trion resonance when holes are added to the monolayer ($X^+$), and the negative trion resonance when electrons are added ($X^-$). Their state compositions are shown in Fig.~\ref{fig:expMoSe2}(c). The emission of the bright exciton in PL disappears as soon as electrons or holes are added, signifying the efficient generation of trions before hot excitons can reach the light cone.

The interesting point we wish to emphasize is that the trion resonances in Figs.~\ref{fig:expMoSe2}(a) and (b) behave similarly in emission and absorption when $|V_g| \lesssim 3$~V, or equivalently, when the charge density is below $n_c \sim 2\times10^{12}$~cm$^{-2}$. At larger densities, the trion resonances show enhanced broadening and their energies redshift in the emission spectrum versus blueshift in the absorption spectrum. In addition, the observed value of $n_c$ is roughly similar when the monolayer is electrostatically doped with electrons or holes. This observation can be explained by the similar effective masses of electrons and holes in these monolayers.  Interestingly, the change in behavior starts at a relatively large charge density rather than gradually developing from the outset (i.e., from $V_g \sim 0\,$V)

\begin{figure}
\centering
\includegraphics[width=8.5cm]{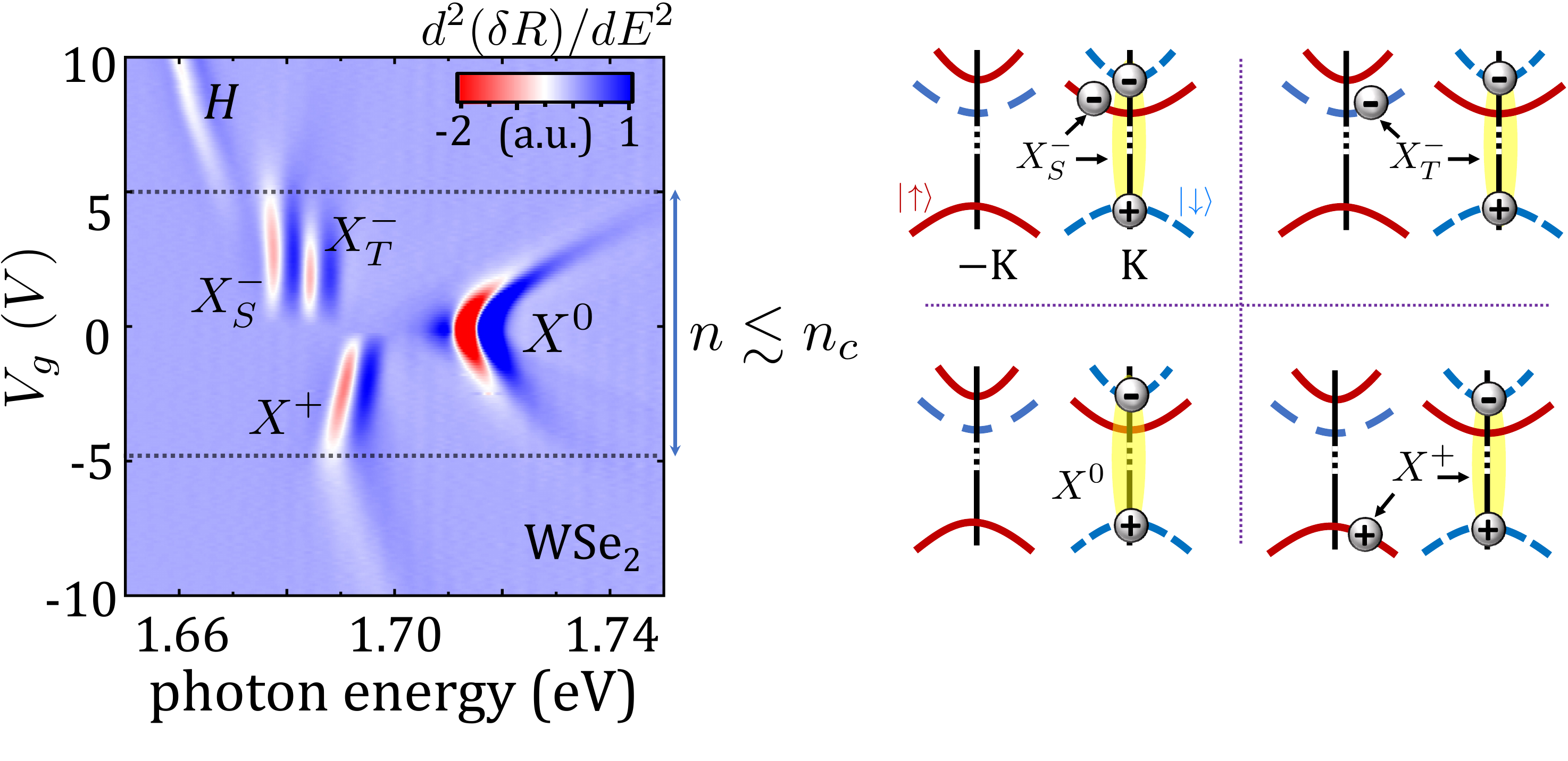}
\caption{ Differential reflectance spectrum of a charge tunable WSe$_2$ monolayer at 15$\,$K \cite{Liu_NatComm21}.  Also shown are the state compositions of bright exciton and trion complexes at relatively small charge densities following photoexcitation of the valley at $K$. The negative trion species are the singlet and triplet ($X_S^-$ and $X_T^-$), depending on whether the two electrons have dissimilar spins or valleys.}\label{fig:expWSe2} 
\end{figure}

The behavior of low-temperature WSe$_2$ monolayers is more subtle, but yet, it shares some commonality with MoSe$_2$ monolayers.  Figure~\ref{fig:expWSe2} shows the low-temperature absorption spectrum of a charge tunable WSe$_2$ monolayer, where changing the gate voltage by 1$\,$V in this device amounts to a change of $\sim4\times10^{11}$~cm$^{-2}$ in  charge density \cite{Liu_NatComm21}. Again, we notice a different behavior between $|V_g |\lesssim 5\,$V and $|V_g |\gtrsim 5\,$V, or equivalently, when the  charge density is below or above $n_c \sim 2\times10^{12}$~cm$^{-2}$. At larger densities, the positive trion resonance starts to significantly broaden and its energy starts to blueshift, whereas the negative trion resonances disappear and a new resonance emerges at lower energies (marked by $H$).  As shown in Fig.~\ref{fig:expWSe2}, the valleys of a  WSe$_2$ monolayer are such that the photoexcited electron belongs to the top spin-split conduction-band valleys, whereas electrostatically-doped electrons populate the bottom spin-split valleys at $\pm K$. As such, the photoexcited pair can be created next to a pair of distinguishable electrons from $K$ and $-K$ without violating the Pauli exclusion principle. The result is a hexciton; a composite excitonic complex with a trion at its center and a satellite electron that is glued to the complex by the two Coulomb holes around the electrons of the trion. Interested readers can find more details in Refs.~\cite{VanTuan_PRL22,VanTuan_PRB22}. 

Here, the point we wish to emphasize is that the switch from trions to hexcitons happens at the same conspicuous density ($\sim n_c$) at which  the positive trion in this compound or positive/negative trions MoSe$_2$ monolayers change their behavior. Interestingly, and possibly not accidentally, the exciton resonances in the absorption spectra of Figs.~\ref{fig:expMoSe2} and \ref{fig:expWSe2} are completely quenched when $n \gtrsim n_c$.  

\subsection{Large charge densities ($n > n_c$)}
 At large charge densities, we can assume a translation-invariant system of weakly-correlated  itinerant electrons,  wherein the wavefunctions of spin-down electrons can overlap with those of spin-up electrons. It is unlikely that screening is causing the onset of enhanced broadening of trions and their energy redshift  in PL vs blueshift in reflectance when $n \sim n_c \sim 2\times10^{12}$~cm$^{-2}$. The simple fact that the hexciton peak ($H$) does not show any signs of broadening, decay or blueshift when $n >n_c$ demonstrates that screening is largely irrelevant. Furthermore, the trions radii in these monolayer semiconductors are $\sim$2-3~nm \cite{Mayers_PRB15,Kylanpaa_PRB15,Kidd_PRB16,Donck_PRB17,Mostaani_PRB17,VanTuan_PRB18}. A similar average distance between particles is reached when the charge density exceeds $10^{13}$~cm$^{-2}$. Unless the density is that large, the charge particles cannot track and screen the Coulomb interactions between the three particles of the small trion \cite{footnote}. 

\begin{figure} 
\centering
\includegraphics[width=8.5cm]{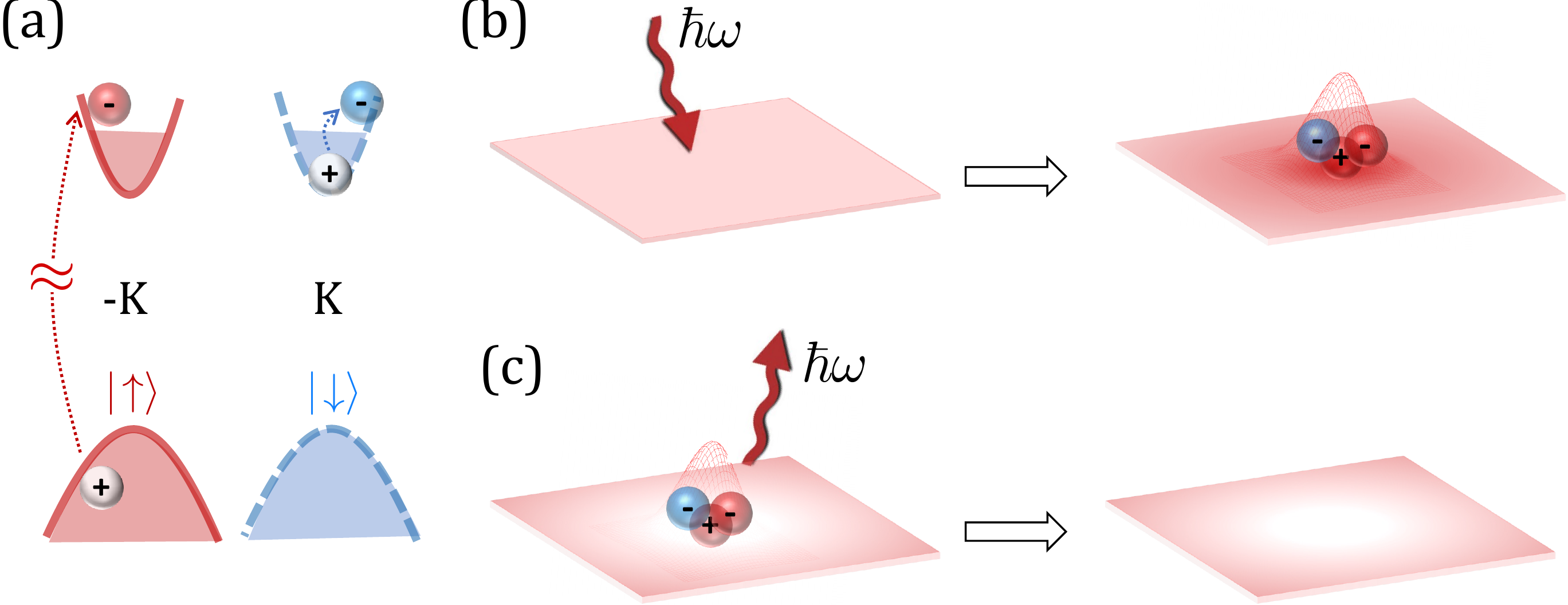}
\caption{ (a) Generation of a trion in electron-doped MoSe$_2$ monolayer following photoexcitation of a spin-up electron in the valley at  $-K$. Creation of the trion involves binding of the photoexcited electron-hole pair in the valley at  $-K$ to a particle-hole excitation of spin-down electron  in the valley at  $K$ (blue sphere). (b) The absorption process starts by photoexcitation of a monolayer with homogeneous charge distribution. The photoexcitation introduces a local surplus of spin-up  electrons, whose density is represented by the redness of the plane. (c) The emission process of a trion  results in a local depletion of spin-up electrons. The spatial distribution of spin-down electrons remains largely homogenous before and after formation (recombination) of the trion in absorption (emission). }\label{fig:absPL} 
\end{figure}

We explain the enhanced broadening of trions and their energy shifts in an electron rich monolayer. Readers can draw similar conclusions for a hole rich monolayer. Figure~\ref{fig:absPL}(a) shows a $k$-space representation of the absorption process in electron-rich MoSe$_2$ monolayer. The photoexcited  electron-hole pair in the valley at $-K$ binds to a spin-down electron following particle-hole excitation in the valley at $K$. The spatial distribution of other spin-down electrons in the valley at $K$ is largely unperturbed. These electrons do not have to get farther away nor come closer to the created trion since they see the same effective charge and subjected to the same Pauli exclusion restriction before and after the trion is created. 

Contrary to the largely unperturbed spin-down electrons, the spatial distribution of spin-up electrons is perturbed by photoexcitation of the valley at $-K$. Figure~\ref{fig:absPL}(b) shows a real-space scheme of the absorption process, wherein spin-up electrons are uniformly distributed in the monolayer prior to photoexcitation. Their density is illustrated by the redness of the plane. The energy blueshift in absorption is a result of the need to use a photon with higher energy in order to create a trion near electrons with the same spin and valley as that of the photoexcited electron. The broadening effect signifies that the lifetime of the final state is ultrafast:  To even out their distribution, spin-up electrons have to scatter out of the congested region, illustrated by the darker red area near the trion in Fig.~\ref{fig:absPL}(b). 

Switching to light emission, the recombination process takes place when the immediate vicinity of the trion is already depleted of spin-up electrons, as illustrated in the left part of Fig.~\ref{fig:absPL}(c). Following photon emission, the charge density of spin-up electrons in the monolayer is inhomogeneous. The emitted photon is bestowed with smaller energy (redshift) because the electron system is left at a higher energy state compared with that of an homogenous gas. The broadening signifies that the lifetime of the post-recombination state is ultrafast, caused by scattering of spin-up electrons into the depleted region.  

\subsection{Small charge densities ($n < n_c$)}
The different behavior when the charge density is below or above $n_c$ in Figs.~\ref{fig:expMoSe2} and \ref{fig:expWSe2}  suggests that the physics changes at this critical density. Otherwise, we would have seen a consistent behavior that develops continuously from vanishing densities. For example, the emergence of the hexciton or enhanced broadening of the trions resonances would start at smaller densities.  

To explain the change in behavior, we conjecture that electrons (or holes) spontaneously break the continuous translation symmetry and start to crystallize when the charge density drops below $n_c$. If Coulomb correlations are relatively strong at low temperatures then the many-particle wavefunction, when viewed as the function of a single coordinate, vanishes at the location of other particles. The number of zeros in the wavefunction increases when electrons (or holes) are added. With that, the number of crests and  troughs of the many-body wavefunction increases. Since the kinetic energy is commensurate with the curvature of the wavefunction, the crystallization breaks at large enough densities. The particles become itinerant and weakly correlated, the translation symmetry of the system is restored, and the wavefunctions of two  particles can now overlap if they have different spin and/or valley degrees of freedom. This regime facilitates the behavior we observe in the optical spectra when $n> n_c$: the trion resonances broaden and behave as discussed in Fig.~\ref{fig:absPL}, the hexciton composite has the right conditions to emerge, and  excitons cease to exist because particles with which the exciton can form a trion or hexciton are `everywhere' in the sample.

As mentioned in the introduction of this paper, diffusion Monte Carlo and Fermionic neural network simulations support the crystallization of charge particles. Using typical ballpark values for MoSe$_2$ monolayer that is encapsulated in hexagonal boron nitride, we find that $r_s \sim 12$ at $n_c \sim 2\times10^{12}$~cm$^{-2}$  (assuming  an effective mass that is half that of a free electron and $\epsilon \sim 3.1$ for the dielectric constant). While diffusion Monte Carlo simulations predict that a Wigner crystal exists when  $r_s \gtrsim 36$ (or equivalently, $n \lesssim 0.1n_c$) \cite{Attaccalite_PRL02}, Fermionic neural network simulations show evident crystalline order already at $r_s=10$ \cite{Cassella_PRL23}.  Finally, extrinsic disorder sources can further help to sustain the localization of charge particles. In this view, $n_c$ in Figs.~\ref{fig:expMoSe2} and \ref{fig:expWSe2} may not necessarily be an intrinsic parameter of these monolayers. 

 
 \subsection{The importance of this work}

 \begin{figure}
 \includegraphics[width=8.5cm]{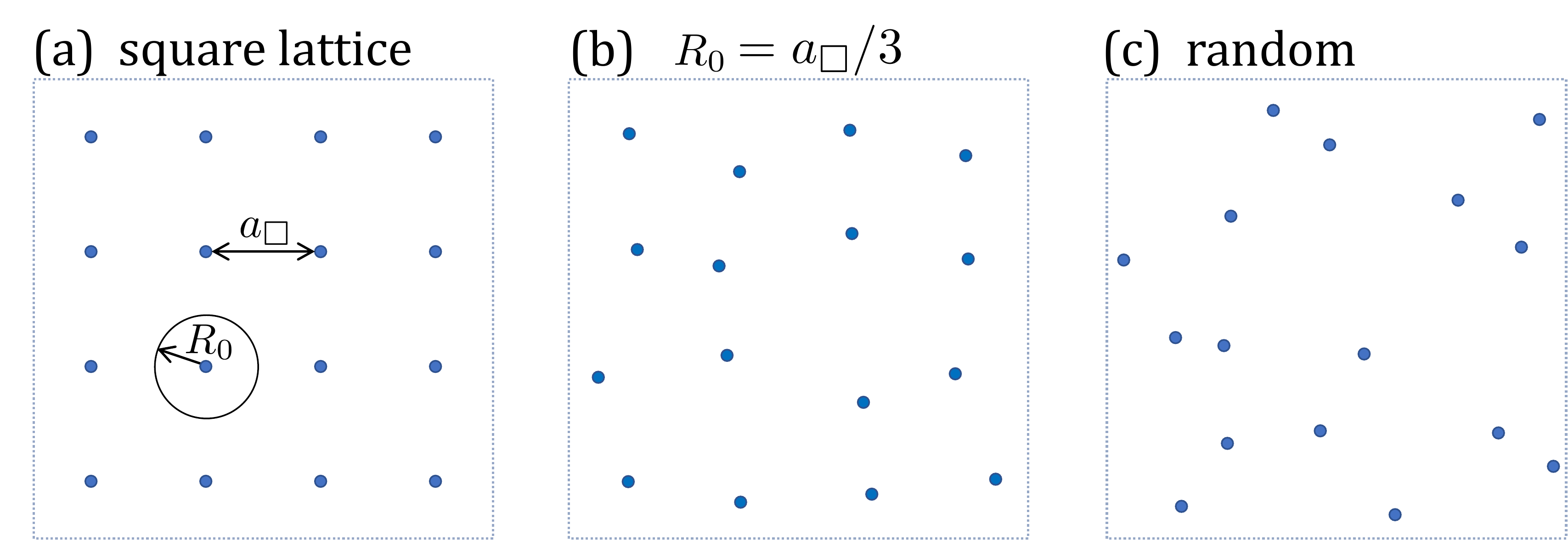}
\caption{ Localized electron configurations, wherein the order is progressively lost from left to right.  (a) Square lattice. (b) A `quasi-ordered' configuration in which the electron is randomly distributed within a circle of radius $R_0=a_\square/3$ around the square lattice site. (c) Random distribution.  }
 \label{fig:sites}
\end{figure}

The  possibility that  charge particles start to crystalize in the monolayer when $n \lesssim n_c$ suggests that the excitonic spectra should be calculated accordingly. We employ the recursion method \cite{Haydock_JPCSSP72}, and investigate the behavior of exciton and trion states in three sample types, as shown in Fig.~\ref{fig:sites}. The first configuration is a Wigner crystal of charge particles, describing a zero-temperature monolayer with small charge density. For simplicity, we study the square lattice case, shown in Fig.~\ref{fig:sites}(a), using both the recursion method and exciton band theory. The results and conclusions of the analysis are qualitatively similar for triangular lattices \cite{VanTuan_arXiv23}. Thermal fluctuations  at nonzero temperatures are mimicked by departure of charge particles from the sites of an ideal square lattice. To do so, the position of each charge particle is randomly distributed within a circle of radius $R_0$ around the square-lattice site. Figure~\ref{fig:sites}(b) shows an example of such `quasi-ordered' distribution when $R_0=a_\square/3$, where $a_\square$ is the lattice constant of the square lattice. The effect of thermal fluctuations for a given $R_0$ is studied by averaging the results over samples with independent quasi-ordered distributions (until convergence is reached). By  repeating this procedure for samples with gradually increasing $R_0$, we can qualitatively emulate the evolution of the optical spectrum when the temperature increases, until the electronic system becomes completely random as shown in Fig.~\ref{fig:sites}(c). 

This work  demonstrates how the energies and broadening of the trion and exciton states depend on charge density, and this dependence is shown for samples with ordered, quasi-ordered or random distributions of charge particles. An important result of this work is to explain how the oscillator strength of trions increases when the charge density increases. The enhanced oscillator strength of trions is demonstrated using two independent approaches. The first one is by calculating the local density of states around an electron (hole) site, showing how the trions borrow the oscillator strength from the crystal around them. The second approach relies on the concept of the giant oscillator strength, originally conceived by Rashba in the late 1950s to explain the very large  oscillator strength of excitons that are weakly bound to impurities \cite{Rashba1957,Rashba1962}. We repeat this derivation for electrostatically-doped monolayers, wherein localized charge particles mimic the role of shallow impurities and trions mimic impurity-exciton complexes.

\subsection{Particle distinguishability} \label{sec:pd}

\begin{figure}
 \includegraphics[width=7.5cm]{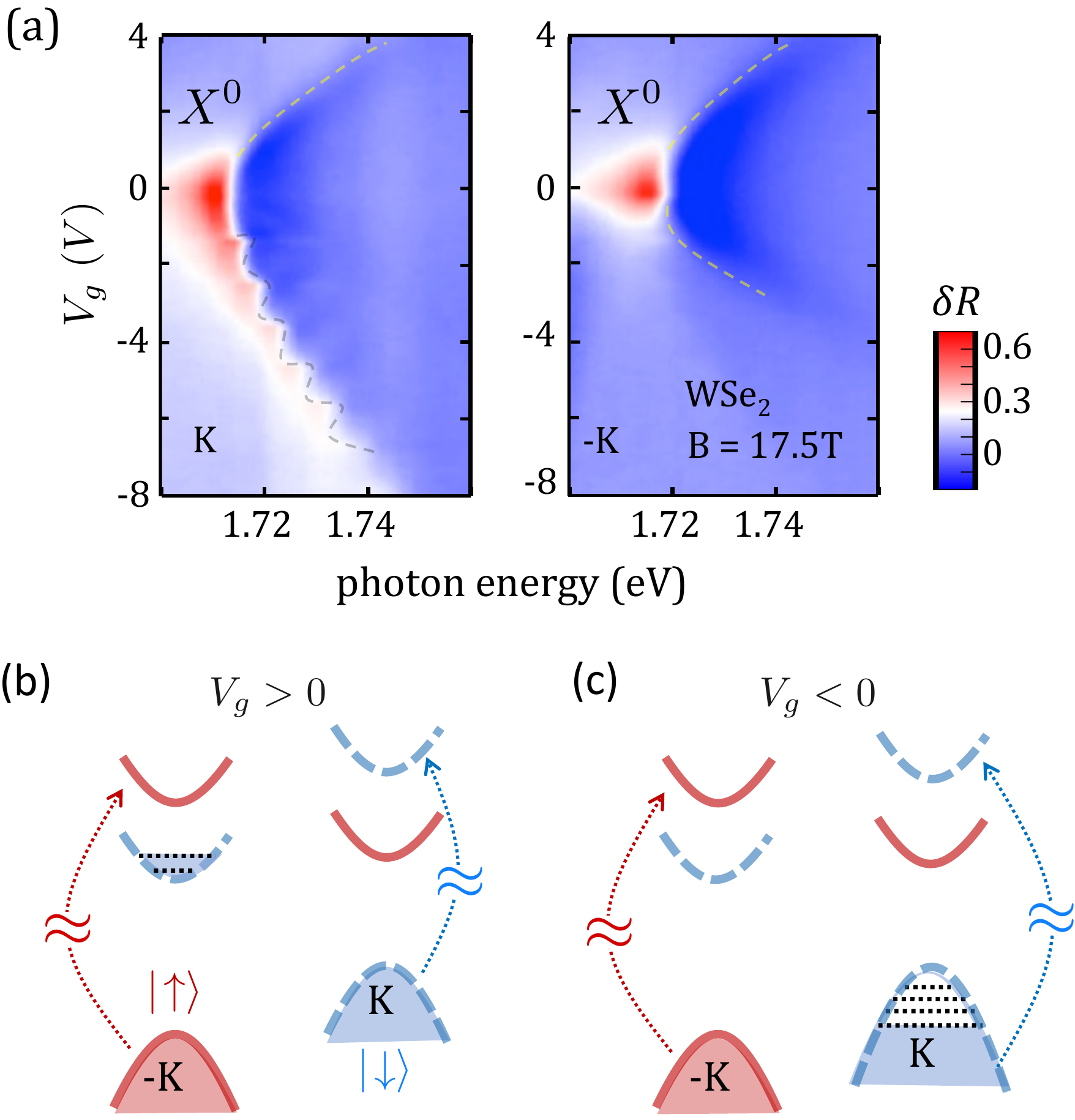}
\caption{Helicity resolved magneto-optical reflectance spectra of the neutral exciton in a charge tunable WSe$_2$ monolayer at 4$\,$K \cite{Liu_PRL20}. The out-of-plane magnetic field is 17.5$\,$T. As a guide to the eye, the faint dashed lines trace the energy blueshift of the exciton. (b) Helicity resolved optical transitions when the electrons in the monolayer are fully spin and valley polarized. (c) Helicity resolved optical transitions when the holes are fully spin and valley polarized. }
 \label{fig:magR}
\end{figure}

Another important aspect that we address in this work is to explain the dependence of exciton energy on the type of electrostatically-doped charge particles in the monolayer. Namely, how the energy blueshift of the exciton depends on whether its electron (hole) has similar or dissimilar spin and valley quantum numbers compared with those of electrons (holes) in the monolayer. To exemplify the importance of particle distinguishability, we use the magneto-absorption spectra of the neutral exciton in WSe$_2$ monolayer, as shown in Fig.~\ref{fig:magR}(a).  These helicity (valley) and density dependent maps were taken at $B$=17.5~T \cite{Liu_PRL20}.  Given the large magnetic field, the electrons are fully polarized in the  bottom conduction-band valley at $-K$  when applying a  small positive gate voltage, as shown in Fig.~\ref{fig:magR}(b). Similarly, the holes  are fully polarized in the top valence-band valley  at $K$ when applying a small negative gate voltage, as shown in Fig.~\ref{fig:magR}(c).  The gate-voltage window of complete valley and spin polarization is larger for holes because of the larger $g$-factor at the top of the valence band \cite{Robert_PRL21}.

Comparing the spectra of Figs.~\ref{fig:expWSe2} and \ref{fig:magR} (without and with a magnetic field), we see  that the energy blueshift of the exciton in all cases is more than 20~meV when the gate voltage increases from 0 to $\sim$4\,V. Namely, the blueshift rate is independent of whether the electrons are distributed evenly between the bottom valleys at $\pm K$ ($B=0$) or fully polarized in one valley ($B=17.5\,$T). This independence stems from particle distinguishability: the photoexcited electron belongs to the top valley whereas electrostatically-doped electrons populate the bottom valleys.  As a result, photoexcitation of the conduction-band top valley at $K$ creates triplet (singlet) trions with electrons from the bottom conduction-band valley at $-K$ ($K$), and vice versa for photoexcitation of the valley at $-K$.  The effect of a strong magnetic field in this case is to control which of the two trion species is formed for valley-specific photoexcitation.

The situation is different in WSe$_2$ monolayer that is electrostatically-doped with fully spin-polarized holes. Positive trions cannot be formed if the photoexcited hole has the same spin and valley as those of the holes in the monolayer. In the example shown in Fig.~\ref{fig:magR}(c), the result is that photoexcitation of the valley at $K$ ($-K$) creates neutral excitons (positive trions).  Examining the behavior in Fig.~\ref{fig:magR}(a) when $V_g < 0$, the energy blueshift of the exciton from $K$ is affected by band filling. This effect is evident from the staircase shape of the signal, wherein the optical transition is progressively blocked by continuous filling of hole Landau levels in the valley at $K$. On the other hand, the exciton from $-K$  is not subjected to any band filling effects (Fig.~\ref{fig:magR}(c)), and  yet, the same blueshift is reached at half the hole density. Figure~\ref{fig:magR}(a) shows that the energy blueshift of the exciton from $-K$ is more than 20~meV when the voltage changes from 0 to nearly -4$\,$V whereas the exciton from $K$ experiences a similar blueshift when the voltage changes from 0 to -8$\,$V.  That the energy blueshift of the exciton is governed by particle distinguishability whereas the  band filling effect plays a secondary role is seemingly a counterintuitive result. We will explain this behavior in this work. 

\section{Theory}\label{sec:theory}

Our goal is to investigate the excitonic spectrum  for various charge densities  in the regime $n \lesssim n_c$ with various order configurations (Fig.~\ref{fig:sites}). To explain how we tackle this problem, we first write the envelope wavefunction of the exciton under the effective mass approximation,
\begin{equation}
\Psi_x ({\bf r}_e,{\bf r}_h)  = \psi(\bm{r}) \phi(\bm{ \rho}).
\label{eq:wavefun}
\end{equation}   
$\bm{\rho} ={\bf r}_e - {\bf r}_h $  is the relative motion coordinate between the electron and hole, and ${\bf r}=(m_e{\bf r}_e+m_h {\bf r}_h)/M_x$ is the translation coordinate (center of mass). $m_e$ and $m_h$ are the electron and hole effective masses, respectively, and $M_x = m_e+m_h$ is the translational mass. Given that the exciton radius, $\langle \rho  \rangle$, is much smaller than the average distance between charge particles  in the monolayer, we can neglect the part that comes from the internal relative motion of the electron and hole components in the exciton, and treat the small exciton complex as one body with translational mass $M_x$. In the intrinsic limit ($n \rightarrow 0$), the translation part of the wavefunction is a plane wave $\psi(\bm{r})= \exp({i\bm{kr}})$. The exciton is free to move in the two-dimensional monolayer, and the resulting density of states (DOS) is a step function $D(E>0) = M_x/\pi \hbar^2$, where the zero energy level is the exciton resonance at the light cone, $E_{k\rightarrow 0} = 0^+$.  

The presence of charge particles changes the simple plane wave picture. The aim of this work is to calculate the resulting DOS function, through which we can understand how trions emerge at $E<0$, how their oscillator strength depends on charge density,  how the energy shift of the exciton depends on charge density, and how quasi-order and random distributions of the charge particles affect the energy shifts and  lead to broadening. 

To study the DOS function, we invoke the recursion method  \cite{Haydock_JPCSSP72}. This method has been successfully employed to study strongly correlated systems \cite{JaklicAdPhys2000, AichhornPRB2003}, high-temperature superconducting states \cite{ElbioRMP1994}, quantum-spin and Hubbard models  \cite{ElbioRMP1994, Koch2011,TharathepPRB2021}, dynamical effects in the context of density functional theory \cite{Koch2011,Koch2008}, and transport properties of disordered graphene systems \cite{VanTuan_PRB2012,VanTuan_NanoLett2013, Aron_AdvM2014,Alex_PRL2013,VanTuan_PRB2016}. In case that the charge particles are ordered in a periodic lattice, the DOS functions can be obtained by either band theory or the recursion method. Verifying that these two independent methods yield identical DOS functions provides reassurance of the results. Below, we first introduce how the DOS functions are calculated by each method for a general problem, and then we apply the recursion method to study excitons in electrostatically-doped semiconductors. 

\subsection{DOS function from band theory}

 The first method employs the energy dispersion $E_n(\bf k)$ of the $n^\text{th}$ band in the exciton band structure obtained  from band theory \cite{VanTuan_arXiv23}. The DOS function is extracted from
\begin{eqnarray}
D(E) &=& \frac{1}{A} \sum_{n,\bf k} \delta \left( E-E_n({\bf k})\right)  \nonumber \\
&=& -\frac{1}{4 \pi^3}\sum_{n} \int   \Im m   \left( \frac{1}{ \varepsilon -E_n({\bf k})  } \right)d{\bf k} \,\,,\,\,   \,\,\,\,   \,\,\,\,  
\label{Eq:BandMethod}
\end{eqnarray} 
where $\varepsilon= E + i \gamma$. The broadening parameter $\gamma$ is introduced to regularize the delta function, and it represents effects due to finite lifetime, interaction with phonons, crystal imperfections,  and other sources of disorder. 

The band theory is applicable when studying periodic systems with well-defined unit cells in which the wavevector $\bm{k}$ is a good quantum number (here, the wavevector of the exciton due to the presence of a Wigner crystal). For disordered systems or if the unit cell includes large number of particles (e.g., $>10^6$), the recursion method  is a useful approach to calculate the DOS function. 
\subsection{DOS function from the recursion method}

Let $H_\text{TB}$ be a sparse and very large matrix that denotes the tight-binding Hamiltonian of the problem in hand. Deriving $H_\text{TB}$ from the exciton's continuum Hamiltonian  is explained in Sec.~\ref{sec:cont} and Appendix~\ref{app:discrete}.   Using the stochastic method of traces \cite{Skilling_book,Weibe_RMP06,Iitaka_PRB04}, the DOS function can be calculated from expectation value of the DOS operator, $\delta(E- H_\text{TB})$, between any vector of the form \cite{Drabold_PRL93}
\begin{equation}
| \varphi_{\text{RP}} \rangle = \frac{1}{\sqrt{M}} \sum_{J=1}^{M} e^{i 2 \pi \theta_J}| \varphi_J \rangle.
\label{Eq:RPS}
\end{equation}
$M$ is the dimension of $H_\text{TB}$ and $\theta_{J\,\in}\,[0,1]$ is a random phase at site $J$. The DOS function can be improved by averaging over $N_{\text{RP}}$ random-phase states \cite{Iitaka_PRB04,Weibe_RMP06},
\begin{equation}
D(E) = \frac{M}{N_{\text{RP}} } \sum_{s=1}^{N_{\text{RP}}} \left\langle \varphi^{s}_{\text{RP}} \left| \delta(E- H_\text{TB})\right| \varphi^{s}_{\text{RP}} \right\rangle,
\label{eq:DOSRe}
\end{equation}
where the statistical error is of the order $1/\sqrt{M N_{\text{RP}}}$.

To avoid diagonalizing the Hamiltonian of large systems with millions of sites, the Lanczos method is used to transform  $H_\text{TB}$ to a tridiagonal matrix. The Lanczos orthonormal  basis is built from the recursive step,
\begin{eqnarray}
 a_n &=& \langle \chi_n\vert H_\text{TB} \vert \chi_n\rangle ,\nonumber \\
 \vert \tilde \chi_{n+1}\rangle &=& H_\text{TB} \vert \chi_n \rangle - a_n \vert \chi_n \rangle -  b_{n-1} \vert \chi_{n-1} \rangle , \nonumber \\
 b_n &=& \sqrt{\langle \tilde \chi_{n+1}\vert \tilde \chi_{n+1}\rangle} ,\nonumber \\
 \vert \chi_{n+1}\rangle &=& \frac{1}{b_n}\vert \tilde \chi_{n+1}\rangle,
 \label{Eq:Recur}
 \end{eqnarray}
for $n \ge 1$, where the base case ($n =0$) is $\vert \chi_1 \rangle= \vert \varphi_{RP} \rangle$ and $b_0=0$.  The recursion coefficients  $a_n$ and $b_n$ are, respectively, diagonal and off-diagonal elements of the Lanczos matrix,
\begin{equation}
H_\text{L} = \begin{pmatrix}
                a_1 & b_1 &        &        &     \\
                b_1 & a_2 & b_2    &        &     \\
                    & b_2 & \ddots & \ddots &     \\
                    &     & \ddots & \ddots & b_{N} \\
                    &     &        & b_{N}    & a_N \\
\end{pmatrix}.
\label{Htridiag}
\end{equation}
The advantage of  the Lanczos method is that it transforms an $M\times M$ sparse matrix  to a much smaller $N \times N$ tridiagonal matrix ($H_\text{TB} \rightarrow H_\text{L}$).   The computational cost for such transformation scales linearly with $M$ instead of $M^3$ of the diagonalization method, rendering the method suitable for studying large systems with various configurations \cite{ Torres2020,Tuan2016}. Calculation of the DOS function in the Lanczos basis is straightforward 
\begin{eqnarray}
\!\!\!\!\langle \varphi_{RP} \vert \delta(E-H_\text{TB}) \vert \varphi_{RP} \rangle &=& \langle \chi_1 \vert \delta(E-H_\text{TB}) \vert \chi_1 \rangle\nonumber\\
 &=& \lim_{\gamma \mapsto 0} -\frac{1}{\pi} \Im m \left(  G_1(E,\gamma) \right),
 \label{Eq:Delta}
\end{eqnarray}
where $ G_1(E,\gamma)$ is written in form of a continued fraction
\begin{eqnarray} \label{fractcontinue}
G_1(E,\gamma) &=& \langle \chi_1 \vert  \frac{1}{E+i\gamma -H_\text{L}} \vert \chi_1 \rangle  \\
&=& \frac{1}{\varepsilon -a_1-\dfrac{b_1^2}{\varepsilon -a_2-\dfrac{b_2^2}{\varepsilon - a_3-\dfrac{b_3^2}{\ddots}}}}  \nonumber\,\,\, . \\ \nonumber
\end{eqnarray}
To compute Eq.(\ref{fractcontinue}), we need to terminate the continued fraction after $N$ recursion steps. Appendix~\ref{app:termination} includes details of the termination process, as well as technical details on how to get rid of the noise signal from the continued fraction. 

\subsection{Application to the exciton problem} \label{sec:cont}

\begin{figure*} 
\centering
\includegraphics*[width=\textwidth]{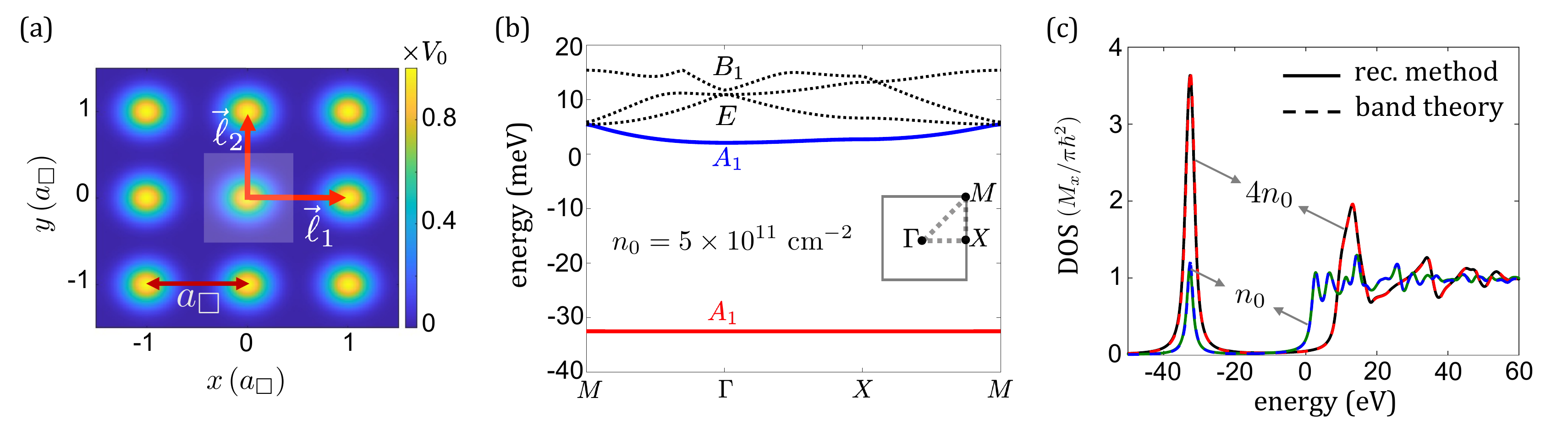}
\caption{ (a) Illustration of the short-range potential profile experienced by an exciton in a square electron lattice. The unit cell and basis vectors $\bm{\ell}_{1,2}$ are highlighted, where the lattice constant is $a_\square = |\bm{\ell}_{1,2}|$. (b) Exciton band structure along axes between high-symmetry points when the charge density is $n_0 = 5 \times 10^{11} \text{cm}^{-2}$.  The energy bands are labeled by the $\Gamma$-point irreducible representations of the square lattice \cite{VanTuan_arXiv23}. (c) Excitonic DOS functions obtained from the recursion method (solid lines) and band theory (dashed lines). The square lattice constants are 14 and 7~nm when the charge densities are $n_0$ and $4n_0$, respectively ($a_\square = 1/\sqrt{n}$). }\label{fig:RecursionVerPseudo} 
\end{figure*}

We use band theory (when applicable) and the recursion method to study the excitonic DOS.  Without loss of generality, we consider a Gaussian function to describe the short-range potential exerted on the charge-neutral exciton by a charge particle. The  potential produced by a landscape of charge particles reads
\begin{equation}
V({\bf r})= \sum_{\ell} V_0 e^{-|{\bf r - r}_\ell|^2/\omega^2}.
\label{Eq:Poten}
\end{equation}  
$w$ is the potential range and $V_0$ is the potential amplitude. We will study how various distributions of potential centers ${\bf r}_\ell$, as shown in Fig.~\ref{fig:sites}, affect the  exciton  spectrum.  

The  recursion method requires to  convert the  continuum Hamiltonian of the point-dipole exciton
\begin{equation}
H({\bf r}) = \frac{\hbar^2 \nabla^2}{2 M_x} + V({\bf r})
\label{Eq:HamilCon}
\end{equation}  
to the tight-binding  Hamiltonian  $H_\text{TB}$. To do so, the 2D sample is discretized to a square grid. The exciton wavefunction is now represented by a set of discrete values  $\{  \varphi_{i,j}  \}$  at the grid points $(i, j)$  along the $x$ and $y$ axes, respectively. By using the nine-point stencil finite-difference formula  for the  Laplacian $\nabla^2$,  the continuum Hamiltonian is recast to (Appendix \ref{app:discrete})
\begin{widetext}
\begin{eqnarray} 
H_\text{TB} = \sum_{ij} \varepsilon_{ij} c^\dagger_{i,j} c_{i,j}\,\, &+& \,\, t  \left( c^\dagger_{i+1,j} c_{i,j} + c^\dagger_{i-1,j} c_{i,j}  + c^\dagger_{i,j+1} c_{i,j} + c^\dagger_{i,j-1} c_{i,j} \right) \,\, \nonumber\\
  &+& \,\,   t'  \left( c^\dagger_{i+1,j+1} c_{i,j} + c^\dagger_{i-1,j-1} c_{i,j} + c^\dagger_{i+1,j-1} c_{i,j} + c^\dagger_{i-1,j+1} c_{i,j} \right).
  \label{Eq:TBH}
\end{eqnarray}
\end{widetext}
The onsite energy is $ \varepsilon_{ij} = V({\bf r}_{i,j})  + 3\varepsilon_0$, where $\varepsilon_0 = \hbar^2 /(2 M_x d^2)$ and $d$ is the grid spacing. The nearest-neighbor and next nearest-neighbor hopping parameters are  $ t = - \varepsilon_0/2$ and $ t' = t/2$, respectively.

\subsection{Parameters}\label{subsec:par}

Unless stated otherwise, we use the following parameters. The exciton translational mass is $M_x = 0.65 m_e$ ($m_e$ is the free electron mass), and the potential parameters are $V_0 = -170$ meV and $\omega = 1$ nm. These are the only three material-specific parameters, and here we choose them to mimic the behavior in TMD monolayers \cite{Yang_PRB22}, where $w$ is comparable to the exciton radius and $V_0$ is chosen to yield a trion-like band  (explained below). The broadening parameter of $\varepsilon= E + i \gamma$ in Eqs.~(\ref{Eq:BandMethod}) and (\ref{fractcontinue}) is $\gamma = 1$~meV.  

The grid (sample) area is $240 \text{ nm} \times 240 \text{ nm} $ with spacing $d=1~\AA$. Out of the resulting $M=5.76\times10^{6}$ grid points, between $\sim$72 and $\sim$1440 are occupied with particles. The corresponding charge densities are between $1.25\times10^{11}$~cm$^{-2}$ and $2.5\times10^{12}$~cm$^{-2}$. The grid points with particles are centers of the short-range Gaussian potential (${\bf r}_\ell$ in Eq.~(\ref{Eq:Poten})).  The $M$ grid points are used to construct the tight-binding Hamiltonian ($H_{\text{TB}}$), which in turn is used to construct the tridiagonal Lanczos matrix $H_{\text{L}}$ in Eq.~(\ref{Htridiag}) through the recursive relations in Eq.~(\ref{Eq:Recur}). The rank of $H_{\text{L}}$  is typically between $N=1000$ and $2000$.  In cases that the charge particles do not form a perfect Wigner crystal (e.g., Figs.~\ref{fig:sites}(b) and (c)), the presented results are calculated by averaging over 100 DOS functions, each with a different charge distribution while keeping all other parameters the same. 


\subsection{Comparing results from band theory and the recursion method  }\label{subsec:comp}

To test the recursion method, we compare its calculated DOS with the one extracted from exciton band theory by assuming a periodic electron configuration. Figure~\ref{fig:RecursionVerPseudo}(a) shows the crystal potential of a square Wigner lattice, along with the unit cell and basis vectors, $\bm{\ell}_{1,2}$. The relation between the lattice constant, basis vectors and charge density is $a_{\square}= |\bm{\ell}_{1,2}| = 1/\sqrt{n}$. Using the pseudopotential method \cite{VanTuan_arXiv23}, the exciton band structure is shown in Fig.~\ref{fig:RecursionVerPseudo}(b) for  $a_{\square} =   14$~nm.  The zero energy reference level is the exciton resonance at $n = 0$. The result of having one site per unit cell (Fig.~\ref{fig:RecursionVerPseudo}(a)), is that only one band has  negative energies, as shown in Fig.~\ref{fig:RecursionVerPseudo}(b). The negative-energy band mimics the trion state, corresponding to an exciton that is tightly bound to a lattice site (electron or hole). The energy of the trion-like band (about $-32$~meV in this example) is mostly governed by the trapping amplitude of the potential, $V_0$, while being weakly dependent on the lattice constant  (charge density). This behavior persists as long as $w  \ll a_\square$, leading to strong localization of trion-like states at lattice sites. The nearly flat nature of this energy band implies very large effective mass, or equivalently,  suppressed hopping between neighboring lattice sites (the exciton stays with the same charge particle). 

The positive energy bands in Fig.~\ref{fig:RecursionVerPseudo}(b) describe states in which the exciton tends to stay away from lattice sites (i.e., not bound to the Wigner crystal).  To check which of the wavefunctions can strongly couple to light, we focus on states near the $\Gamma$-point  (light cone). The reason is that the exciton momentum following light excitation is much smaller than the width of the Brillouin zone, $1/\lambda_x \ll 1/a_\square$, where $\lambda_x$ is the photon wavelength needed to create the bright exciton of the semiconductor.  Among the $\Gamma$-point energy states in Fig.~\ref{fig:RecursionVerPseudo}(b), only the lowest two transform as the identity irreducible representation (IR) $A_1$.  The transformation properties of $A_1$ describe the $1s$-like envelope functions of the trion (ground-state) and lowest exciton state. We are interested in these $s$-type states because the dipole matrix element of their optical transitions does not vanish \cite{Voronov_PSS03,VanTuan_arXiv23}.   

The blue dashed line in Fig.~\ref{fig:RecursionVerPseudo}(c) shows the extracted DOS function when $a_{\square}=   14$~nm, using band theory and Eq.~(\ref{Eq:BandMethod}). The red dashed line shows the corresponding DOS function at $a_\square= 7$~nm. The respective calculations with the recursion method are shown by the solid green and black lines. The perfect agreements between the independent DOS calculations, one with band theory and the other with the recursion method, provides reassurance for both techniques. 

If we ignore for the moment the resonances in Fig.~\ref{fig:RecursionVerPseudo}(c), the DOS is a step function around zero energy. This result is expected for a free particle in two-dimensions (here the exciton).  The resonances that are superimposed on the step function in Fig.~\ref{fig:RecursionVerPseudo}(c) come from the (Wigner) crystal potential.  The trion energy band contributes to the resonance around -32~meV and exciton bands to resonance features in the continuum (positive energies).

\begin{figure*} [t]
\centering
\includegraphics*[width=16.5cm]{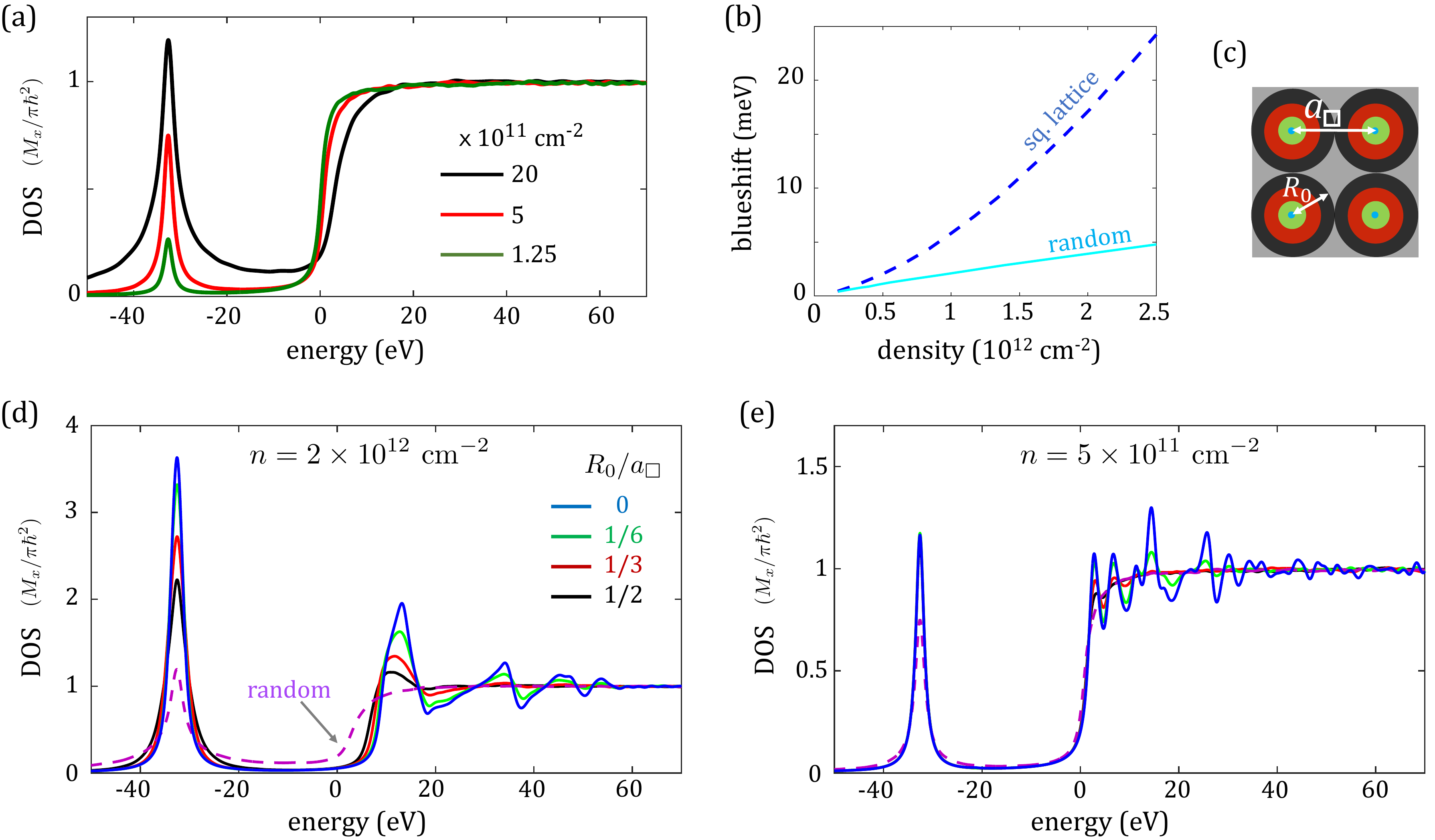}
\caption{ (a) Excitonic DOS functions of various charge densities when the charge particles are randomly distributed. (b) The resulting energy blueshift of the exciton state as a function of charge density. For comparison, the dashed line shows the energy blueshift when the charge particles form a square lattice. (c)  In quasi-ordered configurations, the position of charge particles is randomly distributed within circles of radii $R_0$ around the square lattice sites. (d) and (e) Excitonic DOS functions of quasi-ordered configurations when the charge density is $2 \times 10^{12}\,$cm$^{-2}$ and $5 \times 10^{11}\,$cm$^{-2}$, respectively.}\label{fig:DnO} 
\end{figure*}


Optically active  (bright) excitons contribute to the DOS function near the step region (small positive energies). This result is understood because the optically active exciton belongs to the first energy band in the continuum, as shown in Fig.~\ref{fig:RecursionVerPseudo}(b). The flat nature of this energy band  means that the $\Gamma$-point  coincides with the zero-energy region of the DOS function. Furthermore, regardless of whether the charge particles are ordered or not, the exciton envelop function should be $s$-type in order to enable coupling with light \cite{Voronov_PSS03}. Since the $1s$ state has the lowest energy, we can associate the optically active excitons with states around zero energy  (step region in the DOS) in both ordered and random systems. We will use this property when analyzing simulation results in the next section. 



\section{RESULTS AND DISCUSSIONS} \label{sec:results}

We first analyze the effect of charge density and how the exciton and trion resonances can be used to identify the ordered state of the charge particles. Figure~\ref{fig:DnO}(a) shows the converged excitonic DOS function, calculated by the recursion method after averaging over 100 different random charge distributions. The three DOS functions show that increasing the charge density leads to an increase of the  trion's DOS and energy blueshift of exciton states in the continuum  ($E>0$). The noticeable difference  between the excitonic DOS functions of the square lattice in Fig. \ref{fig:RecursionVerPseudo}(c) and the random distributions in Fig. \ref{fig:DnO}(a) is that resonance features in the continuum are wiped out in the random case. On the other hand, the energy of the trion resonance is kept at $\sim\,- 32$~meV regardless of whether the particles are ordered or randomly distributed. The reason is the strong localization of the trion state around lattice sites, rendering the trion energy less susceptible to a change in the density or distribution of charge particles. The trion energy is kept roughly the same as long as the trion radius is much smaller than the average distance between two charge particles. 

An important difference between ordered and random charge systems is an enhanced energy blueshift and broadening of the exciton in  the ordered case. The broadening can be seen from the smeared step around zero energy. The  blueshift effect can be estimated by looking at the energy at which the DOS function reaches $D_0/2$ at positive energies, where $D_0 = M_x/\pi\hbar^2$ is the exciton's DOS in an intrinsic two-dimensional semiconductor. Figures~\ref{fig:DnO}(b) shows the energy blueshift and how it increases with charge density. For comparison, the dashed line shows the exciton blueshift when the charge particles form a square lattice. The exciton blueshift is evidently weaker when the charge particles lose order. The reason is that a change in charge density has a stronger effect on the packing of particles when they are orderer. This effect is similar when studying Wigner crystals with various symmetries, where a certain change in charge density leads to a larger change in the lattice constant of a Wigner crystal with higher symmetry \cite{VanTuan_arXiv23}.  

To further understand the role of charge order, we study how the excitonic states are affected by a gradual change in the positions of charge particles. Starting with a square Wigner crystal,  the loss of order is introduced by distributing the particles randomly  within a circle of radius $R_0$ around the lattice site of the Wigner crystal, as shown by Fig. \ref{fig:DnO}(c). The higher the value of $R_0$, the less ordered the charge particles are. Figures~\ref{fig:DnO}(d) and (e) shows the excitonic DOS functions for various values of $R_0$ when the charge density is $2 \times 10^{12}$ cm$^{-2}$ and $5 \times 10^{11}$ cm$^{-2}$, respectively.  


The results of Fig.~\ref{fig:DnO} offer three metrics by which the excitonic spectrum can be used to identify when the charge particles become ordered. The first metric is through the enhanced DOS of the trion, which is commensurate with its oscillator strength.  The second and third metrics are through the enhanced energy blueshift and narrowing of the exciton resonance. Comparing the results in Figs.~\ref{fig:DnO}(d) and (e), the three effects are noticeable close to $n_c$ (i.e., when the charge particles are at the verge of becoming itinerant).  These metrics can be tested in experiment by tuning the temperature. For example, consider a monolayer  of MoSe$_2$ or WSe$_2$ with a certain charge density.  The following test can be used to identify the signature of charge order while ruling out the effect of phonons when the temperature is lowered gradually (say from few tens toward 0~K). If suppressed lattice vibrations are evident when the temperature decreases, then changes in the oscillator strengths, energy positions and linewidths of the trion and exciton should be somewhat common to both resonances. On the other hand, if charge order is more relevant, then the narrowing and energy blueshift of the exciton resonance are more evident than those of the trion resonance.

\subsection{ Oscillator strength of trion}

Figures~\ref{fig:RecursionVerPseudo} and \ref{fig:DnO} show that the trion DOS  is significantly increasing when the charge density increases and moderately increasing when the order of the charge particles increases. Below, we provide two independent methods to analyze the oscillator of the trion resonance.  One of the methods will be presented in Sec.~\ref{sec:ldos}, and it relies on the local DOS next to charge particles. The other method, which we present first, relies on the giant oscillator strength concept. This method was originally developed to explain the amplified oscillator strength of excitons that are bound to shallow impurities. Rashba  pointed out that shallow impurity-exciton states are working as antennas borrowing their giant oscillator strength from vast areas of the crystal around them \cite{Rashba1957,Rashba1962}. 

\begin{figure} 
\centering
\includegraphics*[width=8.5cm]{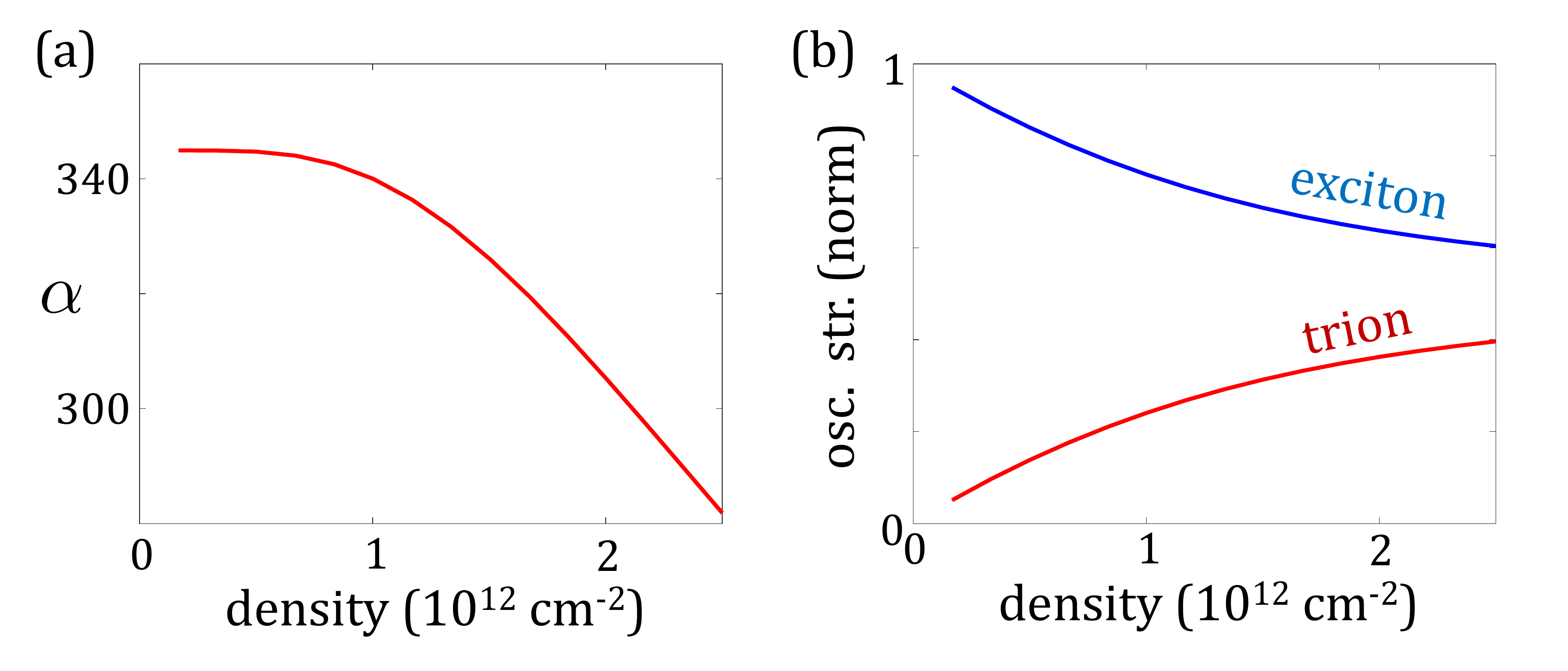}
\caption{ (a) Trion amplification factor, $\alpha=f_T/f_X$,  as a function of charge density, when the charge particles form a triangular lattice. (b) The normalized oscillator strengths, obtained from the relative intensities of the trion and exciton.  }\label{Fig:TrionAmp} 
\end{figure}

\begin{figure*}
\centering
\includegraphics*[width=\textwidth]{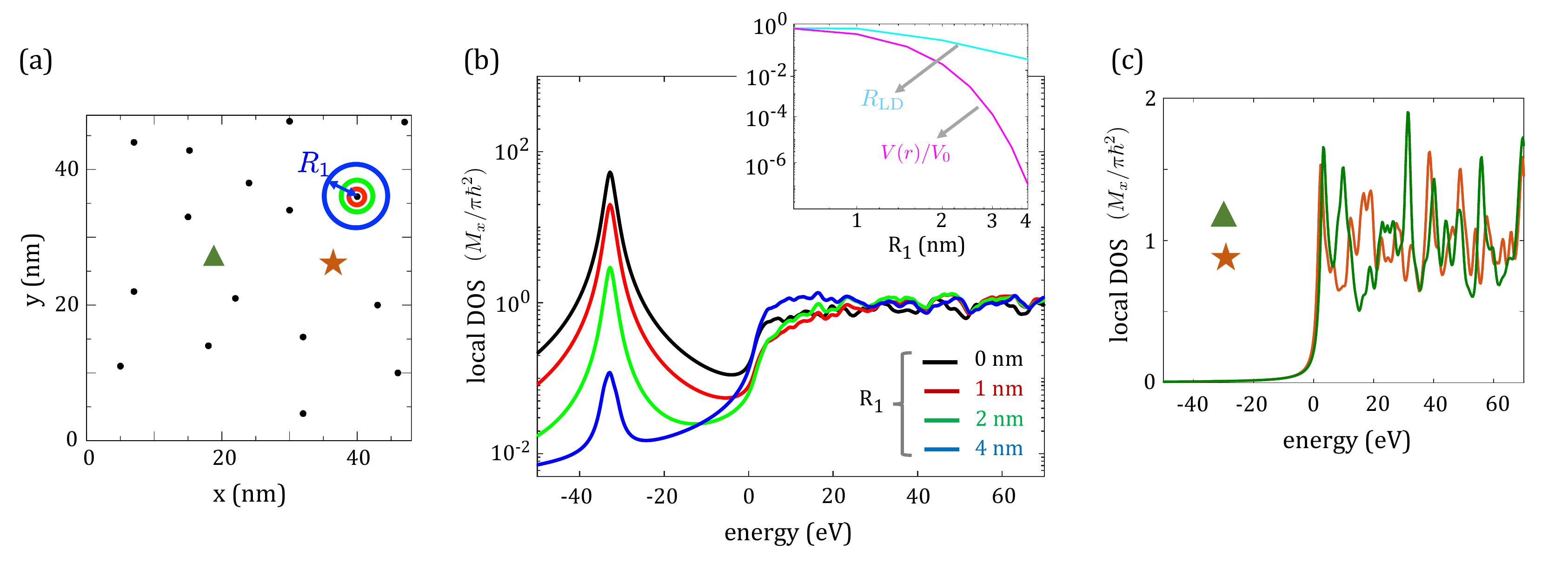}
\caption{ (a) The sample used in calculation of the local DOS function. The sample area is $48\,\text{nm} \times 48 \,\text{nm}$ and the charge density is $n = 7 \times 10^{11}$ cm$^{-2}$. The black dots mark the random positions of charge particles.  (b) The average local DOS function at various distances $R_1$ from the charge particles, as shown by the circles with radii $R_1$ in (a) for one of the particles.  Inset: Decay of the potential and relative trion intensity as a function of the distance from a charge particle. (c) Local DOS function calculations at the positions of the triangle and star marks in (a).
}\label{fig:LDOS} 
\end{figure*}

Here, the role of a shallow impurity is played by a charge particle. The equivalence is reasoned because the gained energy from binding the electron and hole of the exciton is hundreds of meV in TMD monolayers, whereas the gained energy when the charge particle captures the exciton to form a trion is about 30 meV.  Starting with an exciton in the intrinsic semiconductor limit, we let $f_X$ be the oscillator strength contribution  to the exciton resonance by each unit cell of the semiconductor (the atomic unit cell rather than that of the Wigner crystal).  When adding charge particles, the ratio between the oscillator strength of the trion and $f_X$ is given by  \cite{Rashba1962},
\begin{eqnarray} \label{eq:alpha}
\alpha \equiv \frac{f_T}{f_X}   =  \frac{1}{A_c} \left( \int_{A_u} d{\bf r} \,\,\, \psi_T({\bf r}) \right)^2
\end{eqnarray}
where $\alpha$ is the amplification factor, $A_{c(u)}$ is the unit cell of the semiconductor (Wigner) crystal. $\psi_T({\bf r})$ is the envelop wavefunction of the point-like exciton when it is bound to a lattice site (i.e., a trion state). Since the light cone resides near $k=0$, the envelop wavefunction is taken at the $\Gamma$-point of the Wigner crystal. Using band theory, we can write 
 \begin{equation}  \label{eq:psiT}
\psi_{T} ({\bf r})  =  \frac{1}{\sqrt{A_u}}  \sum_{\bf G} u_{T}({\bf G}) \,\, e^{i {\bf G}.{\bf r}}  ,
\end{equation}
where the sum is carried over reciprocal lattice vectors of the Wigner crystal, and the wavefunction has been normalized to its unit cell. Substituting Eq.~(\ref{eq:psiT}) into (\ref{eq:alpha}), we get   \begin{equation}
 \alpha = \frac{A_u}{A_c} u^2_T(G=0).
 \end{equation}
$A_u/A_c$ is the number of unit cells of the semiconductor crystal in one unit cell of the Wigner crystal. $u_T(G=0)$ does not vanish for the $1s$ state at which the light absorption or emission is strongest.  Figure~\ref{Fig:TrionAmp}(a) shows the trion amplification factor $\alpha$ as a function of charge density in a triangular lattice \cite{VanTuan_arXiv23}. The amplification indicates that the trion draws its oscillator strength from about 300 unit cells of the semiconductor crystal in its vicinity.  

The amplification factor can be used to estimate how the relative oscillator strengths of the trion and exciton resonances scale with charge density in absorption experiments. The intensity of the trion scales as $I_\text{T} \sim nA_Lf_T =  nA_L \alpha f_X$, where $nA_L$ in the number of charge particles at the spot area of the light.  The corresponding exciton intensity is $I_\text{X} \sim N_c f_{X} $, where $N_c = A_L/A_c$ is the number of unit cells of the semiconductor crystal under the spot light.  Since we have one charge particle in each unit cell of the Wigner crystal ($nA_u=1$), the relative oscillator strength of the trion is 
 \begin{equation}
 r_\text{T} = 
 \frac{ I_\text{T}}{ I_\text{X}+I_\text{T}} = \frac{u^2_0(0)}{1+ u^2_0(0)}.
  \end{equation} 
The corresponding relative oscillator strength of the exciton is $1 - r_\text{T}$. Figure~\ref{Fig:TrionAmp}(b) shows the relative oscillator strengths of the trion and exciton as a function of charge density. The absorption spectra in Figs.~(\ref{fig:expMoSe2}) and (\ref{fig:expWSe2}) shows that the exciton resonance is largely gone when $n = n_c \sim 2\times10^{12}\,$cm$^{-2}$, whereas the calculated result in Fig.~\ref{Fig:TrionAmp}(b) shows that the oscillator strengths of trion and exciton resonances are comparable when $n \approx n_c$. This apparent discrepancy is settled by the broadening of the exciton resonance. That is, the integrated intensity of the broad exciton resonance in  experiment may indeed be comparable to that of the trion when $n \approx n_c$.

\subsection{Local density of states} \label{sec:ldos}

The fact that trions gain their oscillator strength from areas of the crystal around them is a hallmark of quantum nonlocality \cite{Alain1982, MA2001,Hensen2015, Giustina2015,Shalm2015}. Figure~\ref{Fig:TrionAmp} quantified the oscillator strengths when the charge particles form a Wigner crystal. To calculate the enhanced oscillator strength of trions in case of a random charge distribution, we perform local DOS simulations at various distances from charge particles. As shown in Fig.~\ref{fig:LDOS}(a), the sample area in these simulations is $48 \text{ nm} \times 48 \text{ nm}$ and the charge density is $n = 7 \times 10^{11}$ cm$^{-2}$.  Figure~\ref{fig:LDOS}(b) shows the average local DOS function at a distance $R_1$ from a charge particle (see circles with radii $R_1$ in Fig.~\ref{fig:LDOS}(a)). Regardless of the distance from charge particles, the resonance in the local DOS function emerges at a constant energy. This quantum nonlocality is counterintuitive from the perspective of classical mechanics. Figure~\ref{fig:LDOS}(b) shows that the trion resonance decays when going farther away from charge particles (increasing $R_1$). 
To further confirm this point, we have calculated the local DOS function in two positions that are relatively far from the charge particles, as shown by the star and triangle marks in Fig.~\ref{fig:LDOS}(a). The obtained results in Fig.~\ref{fig:LDOS}(c) show no signature of the trion state in both spectra.  

To quantify the change in the relative intensity of the trion, we evaluate the following ratio from the local DOS function, $D_\text{L}(\text{E})$
\begin{equation} \label{eq:RLD}
R_\text{LD} = \frac{\int_{T} D_\text{L}(\text{E}) d\text{E}}{\int_{TX} D_\text{L}(\text{E}) d\text{E}}.
\end{equation} 
The integration in the numerator is over the trion spectral region, and that in the denominator is over the spectral regions of the trion and exciton. The inset of Fig.~\ref{fig:LDOS}(b) shows the average value of $R_\text{LD} $ as a function of the distance from a charge particle. The integration limits are $[-50,0]$~meV in the numerator of Eq.~(\ref{eq:RLD}) and $[-50,50]$~meV in its denominator. In addition, the inset shows the decay of the short-range potential away from a charge center. Clearly, the local DOS of the trion has longer range. For example, 4~nm from a charge particle, the potential decay is more than five orders of magnitude stronger than that of the local DOS.  That is, the trion collects oscillator strength from regions in which the potential is long suppressed.

\subsection{Particle distinguishability }
So far, the DOS functions were calculated in the presence of distinguishable charge particles in the monolayer. The spin and valley quantum numbers of a distinguishable particle are different than those of the electron and hole in the exciton. The experimental results we have analyzed in Sec.~\ref{sec:pd} have shown that the energy blueshift of the exciton is stronger (weaker) when the  charge particles in the monolayer are distinguishable (indistinguishable). To account for the nature of charge particles in the model, the repulsive interaction between an exciton and indistinguishable particle is simulated by reversing the sign of the short-range potential ($V_0=+170$~meV instead of $-170$~meV that has been used so far for distinguishable particles). The repulsive interaction mimics the role of Pauli exclusion, wherein  antisymmetrization of the wavefunction prevents formation of a trion from an exciton and indistinguishable particle.

\begin{figure} 
\centering
\includegraphics*[width=8cm]{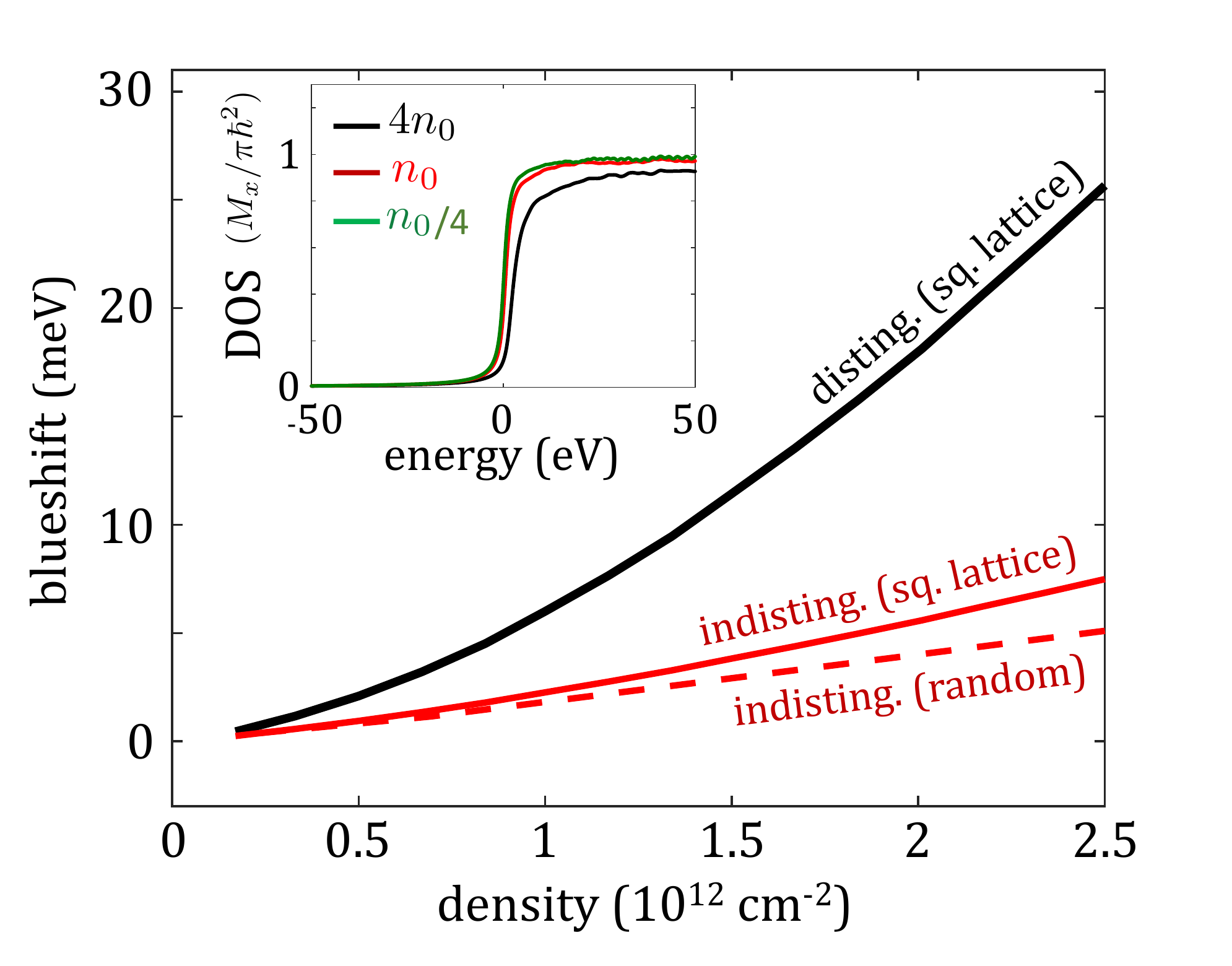}
\caption{Calculated density dependence of the energy  blueshift of excitons when the charge particles are distinguishable and form a square Wigner lattice (black solid line), indistinguishable and form a square Wigner lattice (red solid line), or indistinguishable and randomly distributed (red dashed line). Inset: exciton DOS functions when of the charge particles are indistinguishable and randomly distributed ($n_0 = 5 \times 10^{11} \text{cm}^{-2}$). }\label{Fig:Distin} 
\end{figure}

Figure \ref{Fig:Distin} shows the energy blueshift of the exciton as a function of the charge density of distinguishable particles ($V_0=-170$~meV; black line) and indistinguishable particles ($V_0=+170$~meV; red line). The results are obtained from band theory calculations of a square Wigner lattice, and they support the experimental observation. Namely, the energy blueshift is indeed weaker with indistinguishable particles. Using the recursion method, the dashed line in Fig. \ref{Fig:Distin} shows the energy blueshift  as a function of the charge density of indistinguishable particles that are randomly distributed in the sample. Consistent with previous findings of this work, the energy blueshift of the exciton is enhanced when the charge particles are ordered. The procedure to extract the energy blueshift from the DOS function is the same as the one used in Fig.~\ref{fig:DnO}(b), where the exciton energy follows the blueshift of the step region when the charge density increases. The inset of Fig. \ref{Fig:Distin} shows this behavior through DOS functions of various charge densities. The trion resonance is absent due to particle indistinguishability.  

We explain the difference in energy blueshift of the exciton as  follows.  The ground state of a system with distinguishable particles is the trion whereas the exciton is the excited state. On the other hand,  the ground state of a system with indistinguishable particles is the exciton because trions cannot be formed. The implication of this subtle difference is that in a system with distinguishable particles, the orthogonality between the exciton and trion states  does not permit the exciton to freely choose positions near charge particles wherein the trion state is maximal. In contrast, the exciton in a system with indistinguishable particles has more freedom to minimize its energy by  `finding' the best regions in the sample without having to maintain orthogonality with a lower-energy state. All in all, the energy minimization of the ground state is optimal compared with that of the excited state, leading to slower energy blueshift of the exciton in the presence of  indistinguishable charge particles.

\section{Summary} \label{sec:con}


We have presented an original model to calculate the excitonic behavior in a low-temperature semiconductor. By employing the recursion method, the evolution of exciton states and emergence of trion states is calculated through their density of states when charge particles are gradually added to the semiconductor.  Assuming a two-dimensional semiconductor, the density of states of the exciton is a step function  if no charge particles are present and the exciton is free to move in the semiconductor. When charge particles are added, the excitons sees them as attractive potential centers if their spin and/or valley quantum numbers are distinguishable than those of the electron or hole in the exciton.  The binding of a distinguishable charge particle to the exciton results in the emergence of the trion state at negative energies (the exciton resonance energy in the intrinsic limit is the reference zero-energy level). The theory connects the oscillator strength of the trion to its density of states, which in turn is commensurate with the charge  density of distinguishable particles. 

Several important effects were demonstrated by the theory and corroborated by experimental results. The first effect is enhanced energy blueshift of the exciton state versus a significantly smaller energy shift of the trion state when the charge density of distinguishable particles increases. This behavior cannot be attributed to anti-crossing phenomena of the exciton and trion resonances in which case the energy shifts of the trion and exciton should be similar in magnitude and opposite in sign. The energy of the trion is mostly governed by the amplitude of the short-range potential between an exciton and distinguishable charge particle, while being weakly dependent on charge density if the average distance between nearby particles is larger than the trion radius. In particular, a weak energy shift of the trion resonance is expected when Coulomb correlations cause the charge particles to avoid each other and become localized  or quasi localized.

Compared with the weak energy shift of the trion, the exciton shows a significant energy blueshift in the presence of distinguishable charge particles.  Whereas energy minimization of the ground state (trion) is optimal, the excited state (exciton) has to be both extended and orthogonal with respect to the ground state.  As the charge density of distinguishable particles increases, it becomes harder to satisfy both constraints in conjunction, leading to enhanced energy shift of the exciton. Following the same reasoning, the energy blueshift of the exciton state is relatively smaller when the charge density of indistinguishable particles increases. These particles behave as repulsive potential centers with which the exciton cannot bind and form a trion. The energy blueshift is weaker because the requirement for orthogonality is alleviated (i.e., the exciton is the ground state).  

The theory predicts that a change in charge density affects exciton and trion states the most when the charge particles are optimally packed in the monolayer. For example, the lattice constant of a triangular lattice is modified the most by a change in charge density, that of a square lattice slightly less, and the trend continues when the symmetry and order are gradually degraded \cite{VanTuan_arXiv23}.  By using the recursion technique, we have quantified the changes that excitons and trions experience when the charge particles gradually lose their crystalline phase. Reduced oscillator strength of the trion as well as broadening and slower energy blueshift of the exciton are metrics that can be used to detect the meltdown of the Wigner crystal, especially when the charge density is just below the critical density at which the particles become itinerant. 


\subsection*{Outlook} 

Studying excitons in Wigner crystals through band theory, as we have done in parts of this work and in Ref.~\cite{VanTuan_arXiv23}, is analogous to the theory of Moir\'{e} excitons in van der Waals heterostructures \cite{Yu_SA17,Wu_PRL17}. The potential landscape that point-like Moir\'{e} excitons experience is determined by the twist angle between adjacent monolayers. Excitonic resonances in these heterostructures allow us to identify correlated states in which the Moir\'{e} valleys are populated with charge particles at fractional or integer filling factors \cite{Jin_Nature19,Tang_Nature20,Wang_NatMater23}.  The idea to control the translation motion of excitons through a periodic potential can be extended to systems in which superimposed lattices are patterned by nano-fabrication techniques. This concept can be applied to design THz and (very) far-infrared detectors in which excitons at the ground state are excited to higher-energy states.  The superimposed lattice constant controls the magnitude of energy gaps in the exciton band structure across which optical transitions are allowed by symmetry \cite{VanTuan_arXiv23}.  As such, the spectral window of the detector can be controlled by tuning the periodicity of the fabricated lattice in different regions of the system.

The theory in this work can be applied to study fundamental properties of various electrostatically-doped materials provided the following two restrictions. First, the exciton radius has to be evidently smaller than the average distance between nearby charge particles. As a result, the relatively fast motion between the electron and hole of the exciton can be ignored, and one is left to deal with the translation  coordinate of the point-like exciton (its center of mass motion). The second restriction  is that the charge particles have to be localized or quasi-localized. At low temperatures, the localization can be induced by Coulomb correlations causing the charge particles to crystalize when their density is below a critical density. Above the critical density, the charge particles become itinerant and weakly correlated. Rather than the method used in this work, it is better suited to study excitonic states  in this regime by a theory that accounts for the translation symmetry of the system, wherein the wavevector $\bf{k}$ of a charge particle is a good quantum number. 

Our current understanding is that in semiconductors with tightly bound excitons, the charge particles become itinerant at densities that are still too small to enable  dissociation of the trion by screening. Namely, the charge particles become itinerant, but yet, not fast enough to track the internal relative motion of the three particles in the trion. In transition-metal dichalcogenide monolayers, the trion resonance starts to experience significant broadening when the charge density of itinerant particles continues to increase. The enhanced broadening is introduced because indistinguishable charge particles scatter out of the region in which the trion is created (or into this region in recombination).  This effect can be avoided if we only continue to introduce distinguishable charge particles in the monolayer. It will be interesting to experimentally study the density regime at which the distinguishable trion is dissociated by dynamical screening effects.  
 

\acknowledgments{We deeply thank Chun Hung  (Joshua) Lui and Erfu Liu for providing us with the data in Figs.~(\ref{fig:expMoSe2}), (\ref{fig:expWSe2}) and (\ref{fig:magR}). D. V. T. is supported by the Department of Energy, Basic Energy Sciences, Division of Materials Sciences and Engineering under Award No. DE-SC0014349. H. D. is supported by the Office of Naval Research, under Award No. N000142112448.}


\appendix
\section{Real-space discretization and Tight-Binding Hamiltonian} \label{app:discrete}
 The  exciton wavefunction $| \varphi \rangle $ is discretized at the points $(i, j)$ of a square grid with spacing $d$. The Hamiltonian action on the exciton wavefunction is
\begin{widetext}
\begin{equation}
H({\bf r}) | \varphi \rangle =  \left( - \frac{\hbar^2 \nabla^2}{2M_x} + V({\bf r}) \right) | \varphi \rangle  = V_{ij}\varphi_{i,j} - \varepsilon_0 \left(  \varphi_{i+1,j}+\varphi_{i-1,j} - 2 \varphi_{i,j}  \,\,\,+ \,\,\,   \varphi_{i,j+1}+\varphi_{i,j-1} - 2\varphi_{i,j}   \right)  ,
\end{equation}
where $\varepsilon_0 = \hbar^2 /(2 M_x d^2)$ and $V_{ij} = V({\bf r}_{i,j})$. The TB Hamiltonian becomes
\begin{equation} 
H_\text{TB} = \sum_{ij} \varepsilon_{ij} c^\dagger_{i,j} c_{i,j} + t  \left( c^\dagger_{i+1,j} c_{i,j} + c^\dagger_{i-1,j} c_{i,j} + c^\dagger_{i,j+1} c_{i,j} + c^\dagger_{i,j-1} c_{i,j} \right). 
\end{equation}

\end{widetext}
The onsite energy at  grid point $(i,j)$ is $ \varepsilon_{ij} = V_{ij}  + 4\varepsilon_0 $ and the nearest-neighbor hopping parameter is  $ t = - \varepsilon_0$. The discrete Laplacian $\nabla^2$ in this derivation  is given by convolution with the kernel
 \begin{equation}
 {\bf D}^2_{xy} = \begin{pmatrix}
0 & 1 & 0\\
1 & -4 & 1 \\
0 & 1 & 0
\end{pmatrix}.
 \end{equation} 
 The operator is called the Five-point stencil finite-difference formula and is stable for smoothly varying wavefunctions. However, the formula is not isotropic. To better capture rapidly varying solutions, the nine-point stencil  is a preferred option,
 \begin{equation}
 {\bf D}^2_{xy} = \begin{pmatrix}
\frac{1}{4} \,\,\,\, &  \frac{1}{2}  \,\,\,\,  &  \frac{1}{4} \\
\\
 \frac{1}{2}  \,\,\,\, & -3  \,\,\,\, &  \frac{1}{2} \\
 \\
\frac{1}{4}  \,\,\,\, & \frac{1}{2}  \,\,\,\, &  \frac{1}{4}
\end{pmatrix}.
 \end{equation} 
 Using this operator, one can derive the TB Hamiltonian of Eq.~(\ref{Eq:TBH}).

\section{Termination of the continued fraction} \label{app:termination}
To terminate the continued fraction $G_1$ in Eq.(\ref{fractcontinue}), we define the $n$-th fraction as
\begin{equation}
G_n= \frac{1}{\varepsilon-a_n-\dfrac{b_n^2}{\varepsilon-a_{n+1}-\dfrac{b_{n+1}^2}{\varepsilon- a_{n+2}-\dfrac{b_{n+2}^2}{\ddots}}}}  ,   
\end{equation}
from which we find the recursive step
\begin{equation}
G_n= \frac{1}{\varepsilon  -a_n-b_n^2 G_{n+1}} .
\label{Eq:RecurEq}
\end{equation}
The recursive steps can be applied up to a large value of $n=N$, above which we need to use an approximation to terminate the process. To do so, we make use of the facts that the recursion coefficients $\{a_n, b_n\}$ oscillate around their average values $a$ and $b$, and that the damping of these oscillations is usually fast after a few hundreds recursion steps \cite{Torres2020,Tuan2016}. The approximation we use here is obtained by assigning $G_n =G_N$ for $n>N$ and solve the equation
\begin{equation}
G_N= \frac{1}{\varepsilon -a-b^2 G_{N}} ,
\end{equation}
whose solution is
\begin{equation}
G_N = \frac{(\varepsilon -a) -i \sqrt{(2b)^2 - (\varepsilon -a)^2} }{2b^2}.
\label{Eq:GN}
\end{equation}
Using this approximated solution for $G_N$, we can trace up back and obtain $G_1$ from the recursive steps in Eq.(\ref{Eq:RecurEq}). The DOS functions obtained from the procedure usually contain small noise, which comes from poles of the continued fraction. The noise can be eliminated by employing broadening parameter of a few meV ($\gamma$). However,  the use of large  $\gamma$ can erase real resonances of an ordered charge system.  To deal with this problem,  we not only use a large number of recursion steps $N$, but also average over DOS functions of different $N$. The results in this paper are the average of 100 DOS simulations for $N$ values in the range $1000 \leq N \leq 2000$.

\section{Computational procedure for the recursion method and parameters }
The steps needed to calculate the DOS with recursion method are 
\begin{itemize}
\item Initializing the first vector of the Lanczos basis with  $\vert \chi_1 \rangle= \vert \varphi_{RP} \rangle$, which is the random phase state on each point of the square grid (Eq.~(\ref{Eq:RPS})).
\item  Using the recursive steps in Eq.~(\ref{Eq:Recur})   to calculate the coefficients $\{a_n, b_n\}$. The number of recursion steps $N$ continues until $\{a_n, b_n\}$ converge to constant values $\{a, b\}$.
\item Using Eq.~(\ref{Eq:GN}) to calculate $G_N$ and Eq.~(\ref{Eq:RecurEq}) to trace back to $G_1$.

\end{itemize} 
The DOS is obtain by substituting Eq.~(\ref{Eq:Delta}) into (\ref{eq:DOSRe}), and averaging over $N_{\text{RP}}$ random phase states. Practically, for big samples  wherein $M$ is larger than a few ten millions (the dimension of $H_\text{TB}$), $N_{\text{RP}}=1$ is good enough.  Larger values of $N_{\text{RP}}$ hardly change the DOS. This result is understandable because a big sample is composed of many small samples on which the wavefunctions are random phase states. Finally, we can use multiple configurations to further average the DOS functions in case of dealing with random or quasi-ordered distributions.


\begin{thebibliography}{99}
\bibitem{footnote} To keep the trion bound and small in size, the relative motion of its three particles has to be fast. The relative fast motion makes it harder for electrons or holes in the monolayer to track and screen the Coulomb interaction between the three particles of the trion.
\end{thebibliography}


\end{document}